\documentclass[%
 aip,
 amsmath,amssymb,
 reprint,%
]{revtex4-1}

\usepackage{graphicx}
\usepackage{dcolumn}
\usepackage{bm}

\usepackage[utf8]{inputenc}
\usepackage[T1]{fontenc}
\usepackage{mathptmx}
\usepackage{xcolor}
\begin{document}

\preprint{AIP/123-QED}

\title{Edges of inter-layer synchronization in multilayer networks with time-switching links}

\author{Muhittin Cenk Eser}
\affiliation{Department of Physics, Eastern Mediterranean University, 99628 Famagusta, North Cyprus,via Mersin 10, Turkey}
 
\author{Everton S. Medeiros} \affiliation{Institut f\"ur Theoretische Physik, Technische Universit\"at Berlin, Hardenbergstra\ss e 36, 10623 Berlin, Germany}

\author{Mustafa Riza}
\affiliation{Department of Physics, Eastern Mediterranean University, 99628 Famagusta, North Cyprus,via Mersin 10, Turkey}

\author{Anna Zakharova}
\affiliation{Institut f\"ur Theoretische Physik, Technische Universit\"at Berlin, Hardenbergstra\ss e 36, 10623 Berlin, Germany}

\date{\today}


\begin{abstract}
We investigate the transition to synchronization in a two-layer network with time-switching inter-layer links. We focus on the role of the number of inter-layer links and the time-scale of topological changes. Initially, we observe a smooth transition to complete synchronization for the static inter-layer topology by increasing the number of inter-layer links. Next, for a dynamic topology with the existent inter-layer links randomly changing among identical units in the layers, we observe a significant improvement in the system synchronizability, i.e., the layers synchronize with lower inter-layer connectivity. More interestingly, we find that, for a critical switching-time, the transition to synchronization occurs abruptly as the number of inter-layer links increases. We interpret this phenomenon as the shrinking, and ultimately, the disappearance of the basin of attraction of a desynchronized network state.
\end{abstract}

%
%

\maketitle
%
%
%
%
%
%

{\bf The onset of synchronized behavior is a very intriguing phenomenon in the field of networked dynamical systems. Depending on the connectivity properties of each unit in the network, the transition to synchronization can either occur smoothly with increasing coupling strength between the units or abruptly in a hysteretic scenario. Remarkably, such a variety of transition types has been observed in theoretical and experimental studies across different fields of investigation. More recently, the onset of synchronized behavior has also been addressed in  adaptive networks. In this context, a time-scale for topological changes is prescribed, and the impact on establishing synchronization is assessed. In general, the adaptivity has been found to facilitate synchronization in networks. Here, to gain insight into the types of transition to synchronization triggered by topological adaptivity, we consider a two-layer network of FitzHugh-Nagumo oscillators with a defined time interval (switching-time) between consecutive random inter-layer topological changes. Beyond the expected enhancement in the system synchronizability, we find a switching-time regime where the transition to a synchronized network state occurs abruptly. By analyzing an ensemble of network replicates with different initial conditions close to the transition, we estimate the probability of the desynchronized state as a proxy for its basin of attraction. We attribute the sudden transition to synchronization as the collapse of the basin of attraction of the desynchronized state.}

One of the most important features of spatially distributed systems is the occurrence of common rhythms among their internal interacting elements \cite{Pikovsky2001, Osipov2007}. Such property is well described within the formalism of complex networks via the concept of synchronization \cite{BOC18}. In this context, the occurrence of many different forms of synchronized behavior has been reported in networks across interdisciplinary fields of research \cite{Arenas2008,ZAK20}. Also, methods to assess the stability of synchronized states have been developed both linearly via the system's variational equation \cite{Pecora1998, Pecora2000}, and globally via estimation of the relative size of the system's basin of attraction \cite{Girvan2006,Menck2013} and their basin boundaries \cite{Medeiros2018}. In fact, changes in the basins of attraction of such synchronized states have been found to play a major role in their transitions often occurring in hysteretic scenarios \cite{Zou2014}. 

Transitions to synchronization have been observed in a variety of real-world systems. For instance, in animal physiology, the transition to phase synchronization between cardiac and respiratory oscillations has been observed in rats \cite{Stefanovska2000} and humans \cite{Bartsch2007}. In the brain, the activity of the cerebral cortex has been found to vary from highly synchronous states to desynchronized ones in a critical phenomenon \cite{Fontenele2019}. In technological devices, transitions between synchronization and chaos are induced by resonant tunneling in superlattices \cite{Zhang1996}. Moreover, transition to phase synchronization has been verified between two anharmonic nanomechanical oscillators \cite{Matheny2014}, and between strongly coupled photochemical relaxation oscillators \cite{Cualuguaru2020}. In mathematical models, the transitions to synchronization in networks have been investigated from many different perspectives. A few examples are the different types of transitions to phase synchronization observed in coupled chaotic oscillators \cite{Zhou2002,Osipov2003}, stochastic processes inducing the transition to synchronization in extended systems \cite{Munoz2003}, transitions to synchronization observed in delay-coupled networks \cite{Masoller2011}, the phenomenon of explosive synchronization \cite{Gomez2011,Leyva2012,Zhang2015,JAL19,KUM20,KUM21}, among many others \cite{Boccaletti2016}.

Commonly, the connecting structure of real-world systems is time-dependent, e.g., the brain neuronal plasticity, network of scientific collaborations, or mutation processes in evolutionary biology. To incorporate this feature into a description, many theoretical studies on synchronization have considered the effects of networks with time-varying topologies. Studies show that such settings have massive impact on the transition to synchronization \cite{Gomez2011,Leyva2012,Zhang2015,Boccaletti2016}. In fact, the synchronizability of networks is enhanced for different implementations of the topological time dependencies, i.e., the transitions to synchronization generally occur for lower values of the coupling parameters. The example are provided by, for instance, evolution along commutative graphs \cite{Boccaletti2006}, adaptation of weights \cite{Zhou2006}, on-off coupling \cite{Chen2009}, different rewiring frequencies \cite{Kohar2014}, and random connections \cite{Zhou2019}. More recently, and of particular interest for this study, the enhancement of synchronizability due to time-varying topology has also been shown in networks in which the interacting structure is subdivided into different layers, i.e., {\it multilayer networks}. For instance, in a network of Hindmarsh-Rose neurons in which the coupling structure of both electrical and chemical synapses is time-varying, it has been found that the coupling strength required for the achievement of complete synchrony is significantly lower for increasing the rewiring frequency \cite{Rakshit2018}. Moreover, the enhancement effect has been also observed in a network with multiple layers in which only the structure of specific layers is time-varying \cite{Rakshit2019}. Furthermore, it has been shown that diverse partial synchronization patterns can be generated by multiplexing of adaptive networks \cite{BER20}. Recently, it has been found that noisy modulation of inter-layer coupling strength, called multiplexing noise, induces the inter-layer synchronization of spatio–temporal patterns \cite{VAD20}. Despite such a significant understanding of the role played by time-dependent topologies in facilitating synchronization, some important issues are still await unresolved. In particular, the influence of the topological time-dependencies in the characteristics of the transitions to synchronization is mostly unknown. 
 
In this work, we investigate the transitions to synchronization occurring in a time-switching two-layer network composed of identical FitzHugh-Nagumo oscillators. In isolation, each layer of the system exhibits distinct spatio-temporal patterns due to different initial conditions. By continuously adding connections (links) between mirrored units in the layers, we first observe a smooth transition to a system state exhibiting identical oscillatory patterns in both layers, i.e., inter-layer synchronization. With this procedure, inter-layer synchronization is reached only when every unit of a layer is coupled to its replica node in the other layer, i.e., in the multiplex limit. Next, we show that such a limit is significantly reduced by switching of randomly selected inter-layers links. By changing the time intervals at which the switching is realized, we demonstrate the existence of a regime where the transition to a state with high inter-layer synchronization is critical occurring abruptly for increasing the number of inter-layer links. We explain the mechanism behind such a novel phenomenon as the shrinking, and eventual collapse, of the basin of attraction of the network state with highly desynchronized patterns. This claim is supported by the estimation of the state's basins sizes via their probability of occurrence.

\section{Model}
We investigate the dynamics of coupled FitzHugh-Nagumo (FHN) oscillators. The FitzHugh-Nagumo system is a paradigmatic model originally suggested to characterize the spiking behavior of neurons \cite{FIT61,NAG62}. Synchronization and partial synchronization patterns for this model have been previously investigated in one-layer \cite{Omelchenko2013,ZAK17,Rybalova2019} and two-layer \cite{Mikhaylenko2019,RYB21,SCH21} networks. Moreover, FHN model has been also studied in the context of noise-induced dynamics, e.g., the phenomenon of coherence resonance, for one-layer\cite{MAS17,BAS21} and two-layer \cite{SEM18,MAS21} networks. 

\begin{figure}[!htp]
\centering \includegraphics[height=5cm]{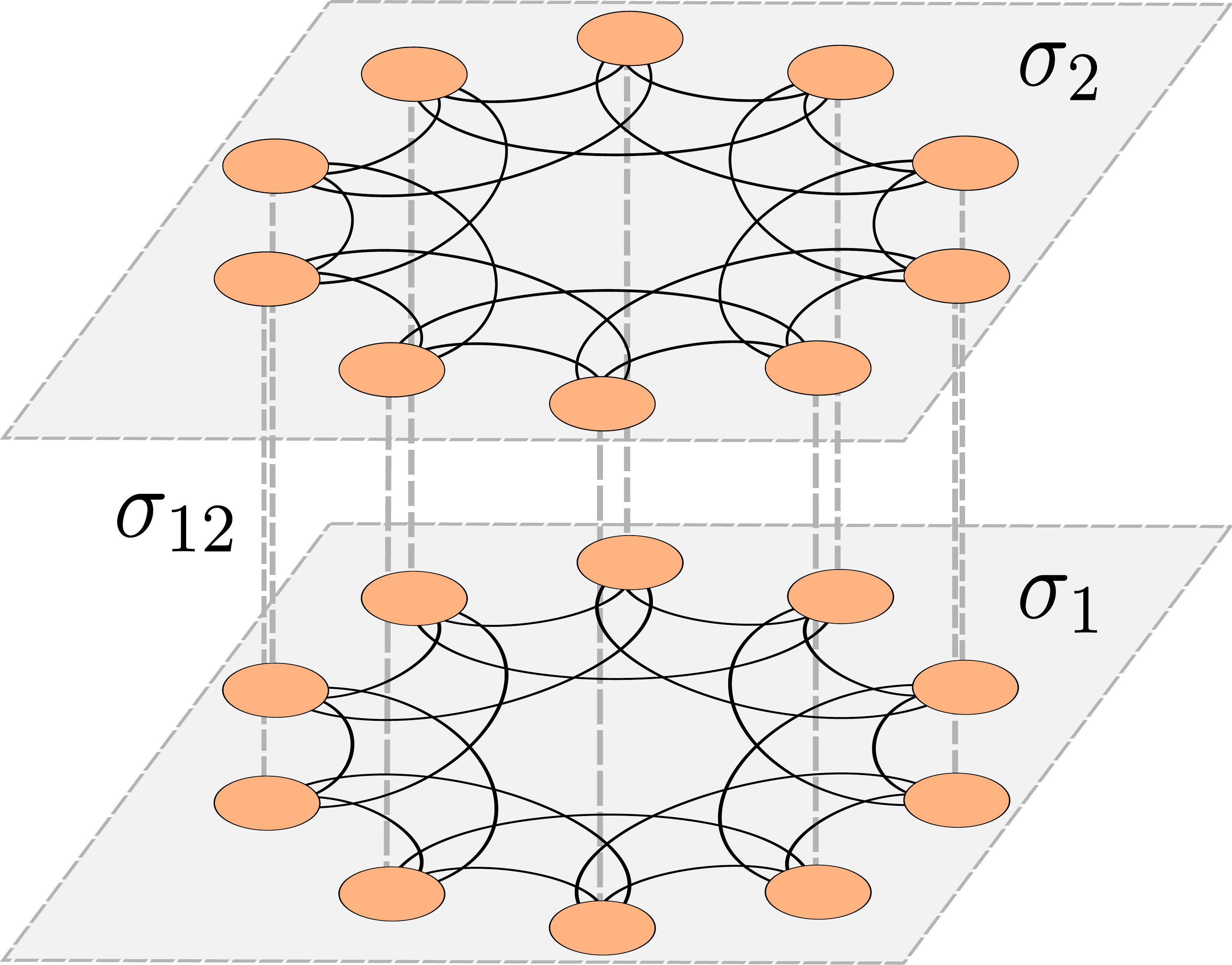}
\caption {Schematic representation of a two-layer multiplex network with nonlocal ring topology within the layers, intra-layer coupling strength $\sigma_1 $,$\sigma_2$ for layer $1$ and layer $2$, receptively, and interlayer coupling strength $\sigma_{12}$.}
\label{figure_1}
\end{figure}

In our analysis, we consider a network consisting of two diffusively coupled layers of interactions, see schematic representation in Fig. \ref{figure_1}. Each layer is composed of $N$ non-locally coupled FHN oscillators. The $2N$-dimensional equation describing the dynamics of the system is given by \cite{Mikhaylenko2019}: 

\begin{equation}
\begin{split}
\varepsilon \frac{du_{1i}}{dt}& =u_{1i}-\frac{{u_{1i}}^3}{3} -v_{1i}+\frac{\sigma_{1}}{2R_1}\sum_{j=i-R_1}^{i+R_1}[b_{uu}(u_{1j}-u_{1i})\\
&\qquad +b_{uv}(v_{1j}-v_{1i})]+\sigma_{12}\left(u_{2i}-u_{1i}\right),\\
\frac{dv_{1i}}{dt}&=u_{1i}+a+\frac{\sigma_{1}}{2R_1}\sum_{j=i-R_1}^{i+R_1}[b_{vu}(u_{1j}-u_{1i})+b_{vv}(v_{1j}-v_{1i})],\\
\varepsilon \frac{du_{2i}}{dt}&=u_{2i}-\frac{{u_{2i}}^3}{3}-v_{2i}+\frac{\sigma_{2}}{2R_2}\sum_{j=i-R_2}^{i+R_2}[b_{uu}(u_{2j}-u_{2i})\\
&\qquad +b_{uv}(v_{2j}-v_{2i})]+\sigma_{12}\left(u_{1i}-u_{2i}\right),\\
\frac{dv_{2i}}{dt}&=u_{2i}+a+\frac{\sigma_{2}}{2R_2}\sum_{j=i-R_2}^{i+R_2}[b_{vu}(u_{2j}-u_{2i})+b_{vv}(v_{2j}-v_{2i})],	
\end{split} \label{eqn:one}
\end{equation}
where $u_{1i}$ and $v_{1i}$ are the activator and inhibitor variables, receptively corresponding to the FHN units populating layer $1$, while $u_{2i}$ and $v_{2i}$ stand for the variables of FHN oscillators populating layer $2$. The FHN oscillators are all identical and rely on two control parameters, namely: $\varepsilon>0$ specifying the time scale separation between fast activator and the slow inhibitor; The excitability threshold $a$ is determining whether the system is in the excitable ($|a|>1$) or oscillatory ($|a|<1$) regime. In this study, we fix the FHN parameters at $\varepsilon=0.05$ and $a=0.5$ (oscillatory regime)  for all numerical simulations. The parameters $\sigma_1$ and $\sigma_2$ specify the intra-layer coupling strength of layer $1$ and layer $2$, respectively. The nonlocal character of the intra-layer coupling is specified by the number of nearest neighbours $R_1$ and $R_2$ of a unit $i$ in layer $1$ and $2$, respectively. A coupling range is defined for each layer by normalizing their respective number of nearest neighbours $R_{1,2}$ with respect to the layer size $N$, i.e., $r_{1,2}=R_{1,2}/N$. The intra-layer coupling function considers cross influence between the activator ($u$) and the inhibitor ($v$), such effect is determined by a rotational coupling matrix \cite{Omelchenko2013}: 
\[
 \begin{pmatrix}
b_{uu} & b_{uv} \\
b_{vu} & b_{vv} 
\end{pmatrix}= \begin{pmatrix}
\cos{\phi} & \sin{\phi} \\
-\sin{\phi} & \cos{\phi}
\end{pmatrix},
\]
where $\phi \in [-\pi,+\pi)$. Depending on a particular choice of $\phi$, a variety of different dynamical behavior can be observed for each layer (isolated). For instance, using $\phi$ as a bifurcation parameter of a single layer system, Rybalova et al. \cite{Rybalova2019} have demonstrated the existence of completely synchronized regimes, solitary states, incoherent regimes, traveling waves, chimera states, and solitary state chimeras. Here, we fix this rotational angle at $\phi=\frac{\pi}{2}-0.1$ which allows the occurrence of chimera states in the intra-layer dynamics. The inter-layer coupling function describes an attractive diffusive coupling scheme with the strength specified by $\sigma_{12}$. In our investigations, the number of FHN units engaged in inter-layer connections is regarded as a control parameter, this quantity is represented by the number of inter-layer links $N_{il}$ with $N_{il} \in [0:N]$. An inter-layer link connects only the two corresponding oscillators of each layer, i.e., oscillators with the same index $i$. Finally, the initial conditions attributed to the system in Eq. \eqref{eqn:one} are randomly distributed outside the region described by equation $r^2-\epsilon\leq u^2+v^2 \leq r^2+\epsilon $. If the nodes are distributed on the circle of radius $r$ or in its close vicinity, the system does not exhibit any chimera states.

\section{Results}

\subsection{Isolated layers}

\begin{figure}[!htp] 
\centering
\includegraphics[width=8.5cm,height=7.1cm]{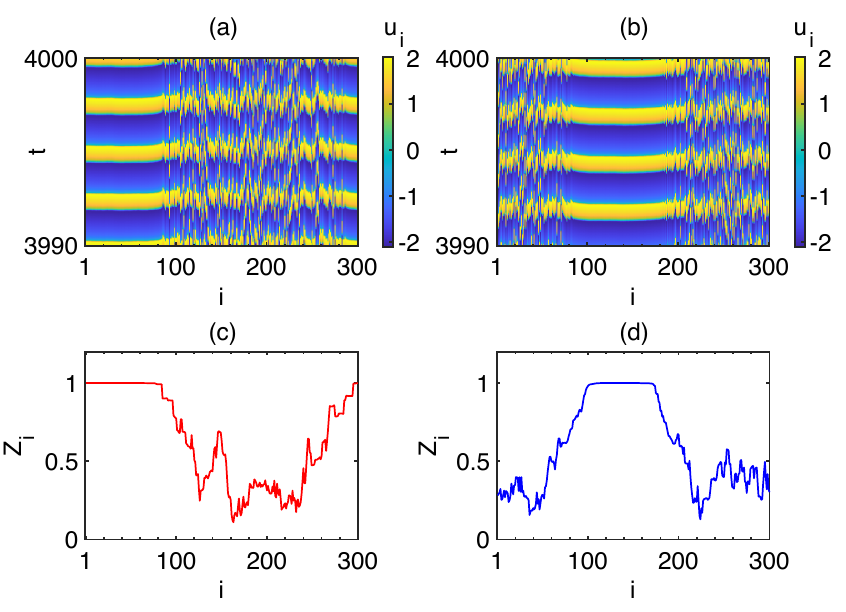}
\caption{Dynamics in the identical isolated layers ($\sigma_{12}=0$) (a) - (b) Space-time plots of the activator $u_i$ for the layers $1$ and $2$, respectively. (c) - (d) The corresponding local order parameter $Z_i$ estimated for every unit $i$ of layer $1$ and layer $2$, respectively. The number of nearest neighbours considered in the estimation of $Z_i$ is $\delta_Z=10$.}
\label{figure_2}
\end{figure}

We start our investigations by first analyzing the layer's dynamics in isolation, i.e., $\sigma_{12}=0$ in Eq. ($1$). We set coupling parameters to be identical in both layers, i.e., $r_1=r_2=0.35$, $\sigma_1=\sigma_2=0.1$ and $N=300$. These parameters are fixed throughout the manuscript. In Fig. \ref{figure_2}(a) and \ref{figure_2}(b), we show the time evolution of all FHN units in layer $1$ and $2$, respectively. Spatially separated domains of coherent and incoherent oscillations are observed in the layers. The coherent domain demonstrates limit-cycle oscillations with the period, in dimensionless units, $T_{lc}\approx2.67$. The coherent domains appear in different spatial locations in the layers due to the choice of layer-specific initial conditions. The coexistence of these two distinct domains of oscillations can be formally classified as chimera states by estimating the local order parameter \cite{Omelchenko2011}:
\begin{eqnarray}
Z_i=\Bigg|\frac{1}{2\delta_Z}\sum_{ \big|j-i\big|\leq\delta_Z} e^{i\Theta_j}\Bigg|,\hspace{1cm}i=1,...,N,
\label{local_order}
\end{eqnarray}
where $\delta_Z$ specifies the number of nearest units considered in the estimation of $Z_i$. The geometric phase of the $j$th element is defined by $\Theta_j=\arctan\left({v_j}/{u_j}\right)$. Coherent domains are indicated by $Z_i=1$, while $Z_i<1$ indicates incoherent ones. In Fig. \ref{figure_2}(c) and \ref{figure_2}(d), we show the local order parameter $Z_i$ for every unit $i$ in layer $1$ and layer $2$, respectively. The occurrence of chimera states in both layers is confirmed by the obtained values of $Z_i$. 

\subsection{Static inter-layer topology}

We now examine the consequences of letting the layers interact via the diffusive coupling, i.e., $\sigma_{12}\neq0$. We have shown in the previous section that in the absence of inter-layer connections, choosing different initial conditions for the layers results in solutions with coherent and incoherent domains at different positions on the ring (see Fig. \ref{figure_2}). Now, in the presence of inter-layer connections, the attractive character of the coupling can naturally lead to the occurrence of inter-layer synchronous behavior, i.e., the spatio-temporal patterns may become identical in both layers. To examine the possible occurrence of such synchronous behavior, we first introduce the global inter-layer synchronization error as \cite{sawicki2018delay}:   
\begin{eqnarray}
E^{12}=\lim_{T\to\infty}\frac{1}{NT}\int_{0}^{T}\sum_{i=1}^{N}\big\|{\bf x}_i^2(t)-{\bf x}_i^1(t)\big\|dt,
\label{inter_sync}
\end{eqnarray}
where ${\bf x}_i(t)=(u_{i}(t),v_{i}(t))$ represents the state space of each FHN unit $i$ with $i=1,\dots,N$. The layers are identified by the superscript index $1$ and $2$. The measure $E^{12}$ is estimated over a time interval specified by $T$. Through this measure, the state of complete inter-layer synchronization is indicated by $E^{12}=0$, while $E^{12}>0$ indicates asynchronous inter-layer behavior. 

\begin{figure}[!htp] 
\centering
\includegraphics[width=9cm]{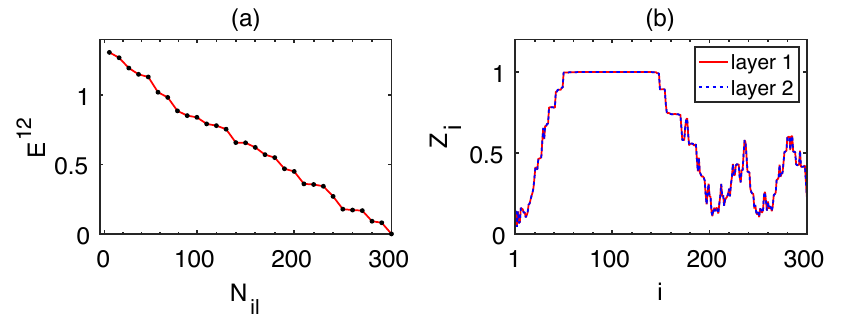}
\caption{(a) The inter-layer synchronization error $E^{12}$ as a function of the number of inter-layer links $N_{il}$. (b) For $N_{il}=300$, the local order parameter $Z_i$ obtained for every oscillator $i$ in the layer $1$ (red dashed curve) and layer $2$ (blue dashed curve). The inter-layer coupling strength is $\sigma_{12}=0.01$.}
\label{figure_3}
\end{figure}

Now, departing from the case of absent inter-layer connections ($N_{il}=0$), the onset of synchronous behavior between the layers can be assessed as inter-layer connections are added. With this, for $30$ different simulations of Eq. ($1$) with the same set of initial conditions, in every simulation, we randomly add $10$ new inter-layer links connecting the $10$ corresponding units of each layer. In other words, the resulting networks contain in the first simulation $10$ randomly distributed inter-layer links, in the second simulation there are $20$ inter-layer connections and so on. In our last simulation, the network reaches the maximum possible number of inter-layer connections ($300$).  Hence, in Fig. \ref{figure_3}(a), for $\sigma_{12}=0.01$, we show the global inter-layer synchronization error $E^{12}$ as a function of the number of inter-layers links $N_{il}$. We observe that the global synchronization error decreases monotonically as the number of inter-layer connections increases. For the case in which all inter-layer links are present, i.e., $N_{il}=N=300$, we observe the inter-layer synchronization of chimera states ($E^{12}\approx0$). To illustrate this fact in Fig. \ref{figure_3}(b), for $N_{il}=N=300$, we show the local order parameter $Z_i$ for the nodes in layer $1$ (red dashed curve) and layer $2$ (blue dashed curve): the spatial location of coherent and incoherent domains of chimera states is the same across the network layers.    

\subsection{Time-switching inter-layer topology}

Next, we study the onset of synchronous behavior between the layers for a non-static inter-layer topology. For this analysis, we consider the trajectory numerically obtained from Eq. ($1$) after a transient of $\tau=3\times 10^3$ in dimensionless units is discarded. Next, in the time interval $t\in[3\times 10^3,4\times 10^3]$, we investigate how the inter-layer synchronization is influenced by switching of the inter-layer links at equally spaced time intervals named switching-times $T_S$. As long as the inter-layer connections are one-to-one between the correspondent oscillators of each layer, at every time moment $T_S$, we re-assign the existent amount of inter-layer links $N_{il}$ across the $N$ available pairs of oscillators.

\begin{figure}[!htp] 
\centering
\includegraphics[width=9cm,height=7cm]{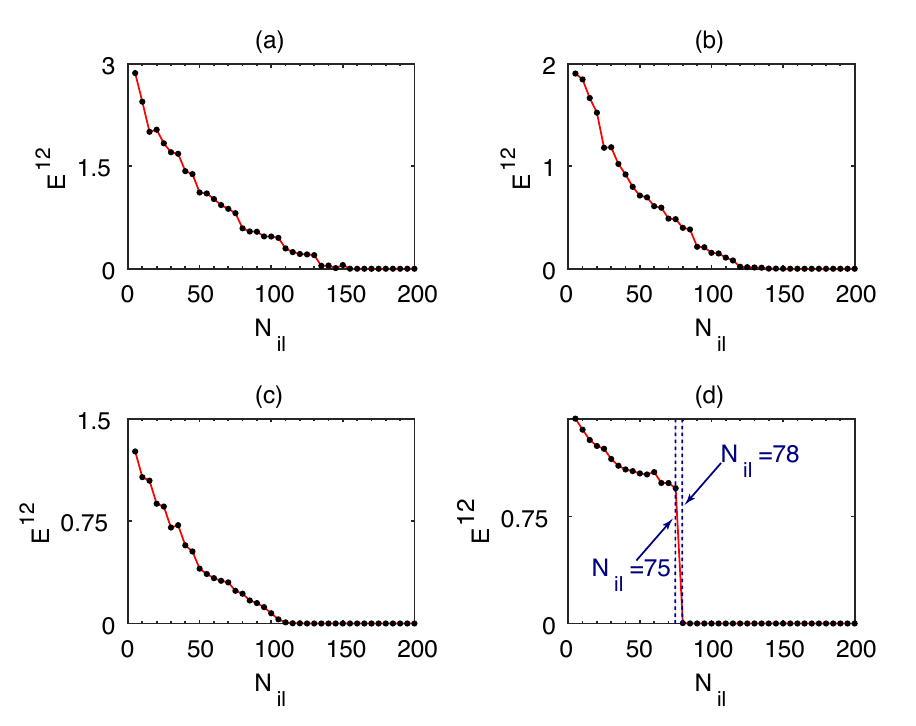}
\caption{The inter-layer synchronization error $E^{12}$ as a function of the number of inter-layer links $N_{il}$ for different switching-times $T_S$. (a) $T_S=100$. (b) $T_S=75$. (c) $T_S=50$. (d) $T_S=23$. The dashed vertical lines in (d) indicate $N_{il}=75$ located before the transition and $N_{il}=78$ located after the transition. The inter-layer coupling strength is $\sigma_{12}=0.01$.}
\label{figure_4}
\end{figure}

With this, in Fig. \ref{figure_4}, we again obtain the inter-layer synchronization error $E^{12}$ as a function of the number of inter-layer links $N_{il}$, but this time for different switching-times $T_S$. In Fig. \ref{figure_4}(a), we show this dependence for $T_S=100$. The system state with synchronous inter-layer behavior is smoothly established as the number of inter-layer links is increased. Similar behavior is observed in Fig. \ref{figure_4}(b) for $T_S=75$, however, the synchronous inter-layer behavior is achieved for a smaller amount of inter-links. For $T_S=50$, in Fig. \ref{figure_4}(c), the synchronous behavior is achieved even for a lower value of $N_{il}$. Interestingly, for $T_S=23$ the transition to the synchronous inter-layer state is abrupt, i.e., after including one additional inter-layer link, synchronization is suddenly established between the layers (Fig. \ref{figure_4}(d)).         

\begin{figure}[!htp] 
\centering
\includegraphics[width=8.4cm,height=7.1cm]{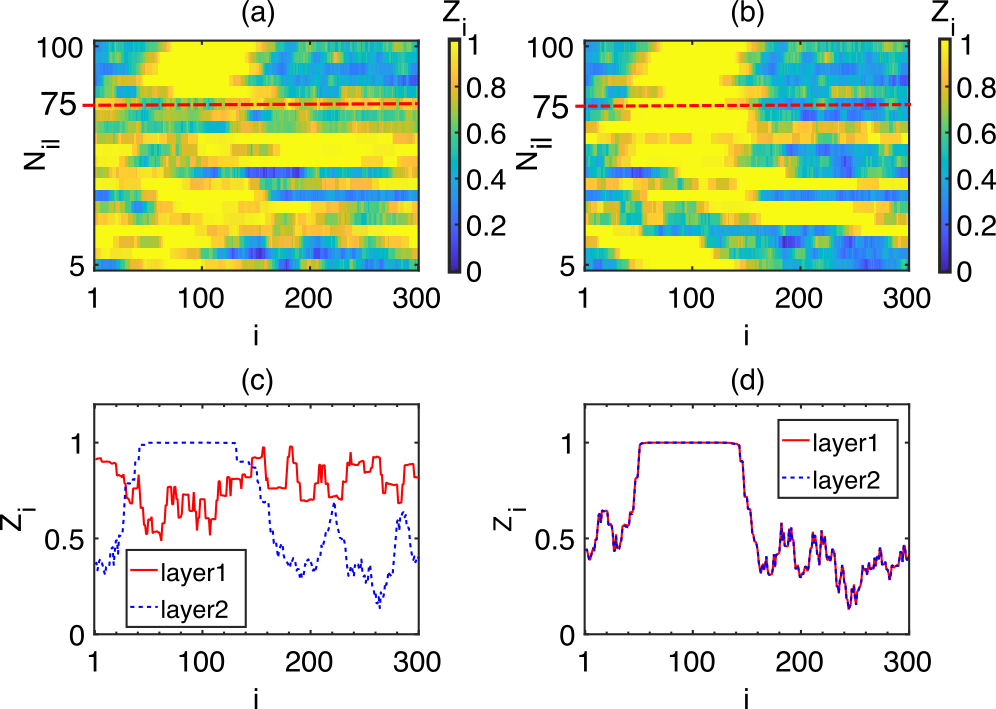}
\caption{For the switching-time fixed at $T_S=23$, in (a)-(b), we show the local order parameter $Z_i$ as a function of the number of inter-layer links $N_{il}$ across layer $1$ and layer $2$, respectively. (c) For $N_{il}=75$, the blue dashed curve indicates $Z_i$ of layer $1$ and the red dashed curve corresponds to $Z_i$ of layer $2$. (d) For $N_{il}=78$ both curves overlap indicating inter-layer synchronization. The number of nearest neighbours considered in the estimation of $Z_i$ is $\delta_Z=10$. The inter-layer coupling strength is $\sigma_{12}=0.01$. }
\label{figure_5}
\end{figure}

Next, we study in more detail the synchronization patterns occurring within each layer. For that, we use the local order parameter $Z_i$ defined in Eq. (\ref{local_order}). This measure quantifies the level of synchronization within the layers by estimating the average deviation of an oscillator $i$ from its neighborhood specified by $\delta_Z$. With this, for $T_S=23$, in Fig. \ref{figure_5}(a) and \ref{figure_5}(b), we obtain the local order parameter $Z_i$ as a function of $i$ and $N_{il}$ for layer $1$ and $2$, respectively. We observe that the layers exhibit different patterns for the values of $N_{il}$ lower than the critical one at $N_{il}=76$. However, for $N_{il}>76$, the patterns in both layers overlap indicating the transition to the synchronized regime between the layers. In addition, we further examine the transition by considering the values of $N_{il}$ before and after the transition (see vertical lines in Fig. \ref{figure_4}(d)). Hence, on one hand, in Fig. \ref{figure_5}(c), for $N_{il}=75$ located right before the transition, the internal dynamics in the layers is significantly different. The blue curve in Fig. \ref{figure_5}(c) indicating $Z_i$ of every oscillator in layer $2$, shows a spatio-temporal pattern containing a coherent and an incoherent domain, i.e., a chimera state; The red curve representing $Z_i$ for layer $1$, discloses an incoherent regime in that layer. On the other hand, in Fig. \ref{figure_5}(d) for $N_{il}=78$ right after the transition, we show that both blue and red curves coincide indicating the same oscillatory pattern (chimera state) in both layers.

\begin{figure}[!htp] 
\centering
\includegraphics[width=8.7cm,height=7.1cm]{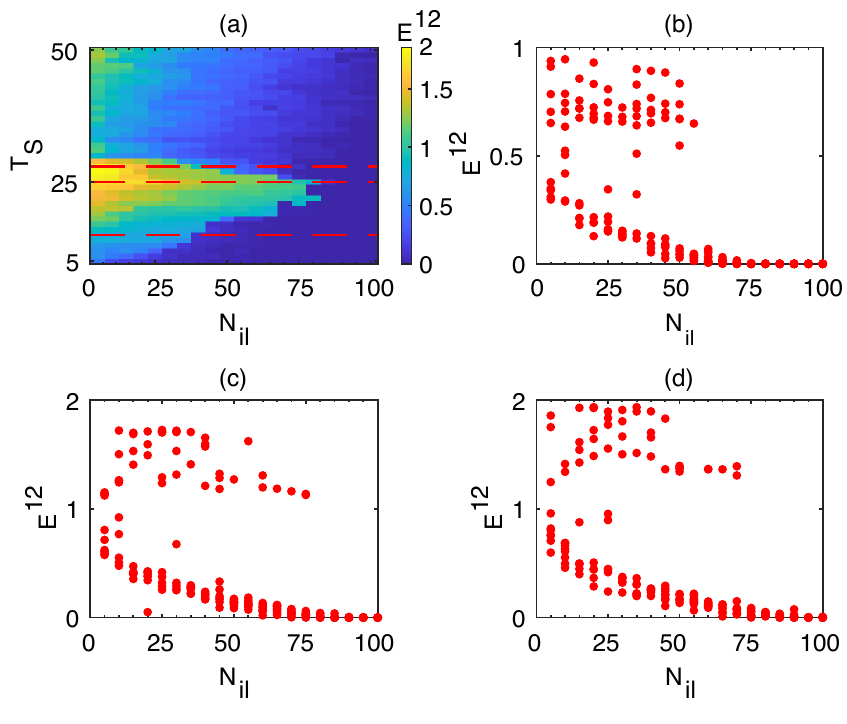}
\caption{(a) Two-dimensional parameter diagram showing the inter-layer synchronization error $E^{12}$ as a function of the number of inter-layer links $N_{il}$ and the switching-times $T_S$. The horizontal lines indicate one-dimensional cuts for specific values of $T_S$ for which the error $E^{12}$ is obtained as a function of $N_{il}$ for different ICs. The respective diagrams are shown in: (b) For $T_S=15$. (c) For $T_S=25$. (d) For $T_S=28$. The inter-layer coupling strength is $\sigma_{12}=0.01$.}
\label{figure_6}
\end{figure}

To investigate the criticality of the transition observed in Fig. \ref{figure_4}(d), we look for possible combinations of the switching-time $T_S$ and the number of inter-layer links $N_{il}$ for which such non-trivial transition occurs. Hence, the diagram in Fig. \ref{figure_6}(a) shows the inter-layer synchronization error $E^{12}$ for different values of $N_{il}$ and $T_S$. In this diagram, the large blue shaded region corresponds to parameter combinations leading to the synchronous inter-layer behavior, while the yellow-green shaded region corresponds to parameters leading to asynchronous behavior. The abrupt transitions under investigation are located at the border between these two parameter regions. Both synchronous and asynchronous states can coexist in the system's state space for the same combinations of $T_S$ and $N_{il}$ depending exclusively on the choice of initial conditions (ICs). To clarify the role of ICs, we choose three different values of $T_S$ (horizontal lines in Fig. \ref{figure_6}(a)) and obtain the synchronization error as a function of $N_{il}$ for $10$ network realizations with different ICs. For $T_S=15$, in Fig. \ref{figure_6}(b), we indeed observe two coexisting regimes for the synchronization error $E^{12}$ that are accessible for different values of the system ICs. The upper branch of the diagram in Fig. \ref{figure_6}(b) indicates an asynchronous state with high synchronization error (HSE), while the lower branch corresponds to a network state at which the layers oscillate with a higher degree of synchronization indicated by low synchronization error (LSE). Similar behavior is observed for the other two horizontal cuts $T_S=25$ and $T_S=28$ shown in Figs. \ref{figure_6}(c) and \ref{figure_6}(d), respectively.   

\begin{figure}[!htp] 
\centering
\includegraphics[width=8.6cm,height=7.1cm]{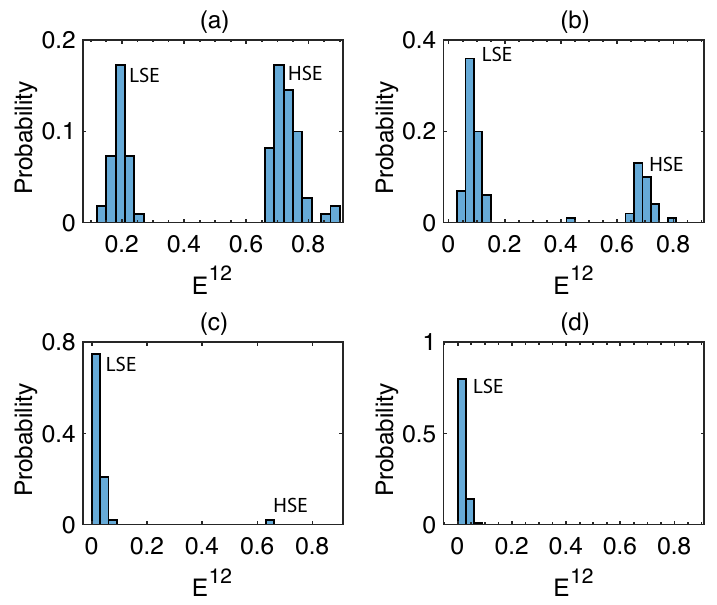}
\caption{The normalized probability of occurrence for each network state, i.e., one with low synchronization error (LSE) and the other with high synchronization error (HSE). The switching-time is fixed at $T_S=15$, the number of inter-layer links (vertical cuts in Fig. \ref{figure_6}(b)): (a) $N_{il}=20$. (b) $N_{il}=40$. (c) $N_{il}=60$. (d) $N_{il}=80$. The inter-layer coupling strength is $\sigma_{12}=0.01$.}
\label{figure_7}
\end{figure}

Furthermore, to unveil the mechanism at which the layers suddenly reach the synchronized regime, we focus on the behavior observed in the diagrams $E^{12} vs. N_{il}$ shown in Fig. \ref{figure_6}(b)-(c) indicating that the system's state-space is indeed shared by two states with distinct synchronization patterns for $N_{il}$ lower than a critical value. Consequently, such coexistence of states implies the partitioning of system's initial state-space in different basins of attraction, i.e., sets of ICs from which the corresponding trajectories approach a determined state. We argue that as $N_{il}$ is increased towards the critical value, the domain of attraction of the state with higher synchronization error (HSE) is gradually shrinking and vanishes at the critical $N_{il}$. To demonstrate that, we estimate the relative size of the domain of attraction for both states, HSE and LSE, via their probability of occurrence. For $100$ network realizations with different ICs, we record the resulting inter-layer synchronization error $E^{12}$ for different values of $N_{il}$. Hence, in Fig. \ref{figure_7}, we show the normalized probability density of the obtained values of $E^{12}$. For $N_{il}=20$, shown in Fig. \ref{figure_7}(a), both states occur with similar probability, indicating that their respective basins of attraction have equivalent size. However, for $N_{il}=40$ shown in Fig. \ref{figure_7}(b), the probability of network trajectories approaching the LSE state is significantly higher indicating the shrinking of the domain of attraction of the HSE state. This tendency is even more evident for $N_{il}=60$ shown in Fig. \ref{figure_7}(c), suggesting that only remnants of the basin of attraction of the HSE state are present. Finally, in Fig. \ref{figure_7}(d) for $N_{il}=80$, all generated trajectories approach the LSE state indicating the complete annihilation of HSE state.             

\section{Conclusions}

In summary, we have investigated the transitions to inter-layer synchronization in a two-layer network composed of identical FitzHugh-Nagumo oscillators. For a static realization of the inter-layer topology, we observe a monotonic dependence of the inter-layer synchronization level with the number of inter-layer links connecting units across the layers. For a dynamic inter-layer topology where a certain amount of inter-layer links switches at prescribed time moments (switching-times), we find a significant reduction of the minimum number of inter-layer links required for the achievement of synchronization between the layers. For particular values of switching-times, we uncover the occurrence of an abrupt transition to inter-layer synchronization for a specific number of inter-layer links. Furthermore, we show that the mechanism behind this critical transition is related to the shrinking and eventual disappearance of the basin of attraction of the desynchronized network state.         

\begin{acknowledgments}
This work was supported by the Deutsche Forschungsgemeinschaft (DFG, German Research Foundation) - Project No. 163436311 - SFB 910.
\end{acknowledgments}

\nocite{*}
\bibliography{multiplex_ref}

\end{document}